\def\Hbar{\overline {\rm H}}
\def\gbar{\overline {\rm g}}
\def\pbar{\overline {\rm p}}

\documentstyle[12pt,epsfig]{article}

\catcode`\@=11
\long\def\@makefntext#1{
\protect\noindent \hbox to 3.2pt {\hskip-.9pt
$^{{\ninerm\@thefnmark}}$\hfil}#1\hfill}		

 \def\@makefnmark{\hbox to 0pt{$^{\@thefnmark}$\hss}}  

\def\ps@myheadings{\let\@mkboth\@gobbletwo
\def\@oddhead{\hbox{}
\rightmark\hfil\ninerm\thepage}
\def\@oddfoot{}\def\@evenhead{\ninerm\thepage\hfil
\leftmark\hbox{}}\def\@evenfoot{}
\def\sectionmark##1{}\def\subsectionmark##1{}}

\newcounter{sectionc}\newcounter{subsectionc}\newcounter{subsubsectionc}
\renewcommand{\section}[1] {\vspace{0.6cm}\addtocounter{sectionc}{1}
\setcounter{subsectionc}{0}\setcounter{subsubsectionc}{0}\noindent
	{\bf\thesectionc. #1}\par\vspace{0.4cm}}
\renewcommand{\subsection}[1] {\vspace{0.6cm}\addtocounter{subsectionc}{1}
	\setcounter{subsubsectionc}{0}\noindent
	{\it\thesectionc.\thesubsectionc. #1}\par\vspace{0.4cm}}
\renewcommand{\subsubsection}[1]
{\vspace{0.6cm}\addtocounter{subsubsectionc}{1}
	\noindent {\rm\thesectionc.\thesubsectionc.\thesubsubsectionc.
	#1}\par\vspace{0.4cm}}

\newcounter{appendixc}
\newcounter{subappendixc}[appendixc]
\newcounter{subsubappendixc}[subappendixc]

\renewcommand{\appendix}[1] {\vspace{0.6cm}
        \refstepcounter{appendixc}
        \setcounter{figure}{0}
        \setcounter{table}{0}
        \setcounter{equation}{0}
        \renewcommand{\thefigure}{\Alph{appendixc}.\arabic{figure}}
        \renewcommand{\thetable}{\Alph{appendixc}.\arabic{table}}
        \renewcommand{\theappendixc}{\Alph{appendixc}}
        \renewcommand{\theequation}{\Alph{appendixc}.\arabic{equation}}
        \noindent{\bf Appendix \theappendixc #1}\par\vspace{0.4cm}}

\def\abstracts#1{{
	\centering{\begin{minipage}{30pc}\tenrm\baselineskip=12pt\noindent
	\centerline{\tenrm ABSTRACT}\vspace{0.3cm}
	\parindent=0pt #1
	\end{minipage}}\par}}

\renewenvironment{thebibliography}[1]
	{\begin{list}{\arabic{enumi}.}
	{\usecounter{enumi}\setlength{\parsep}{0pt}
\setlength{\leftmargin 1.25cm}{\rightmargin 0pt}
	 \setlength{\itemsep}{0pt} \settowidth
	{\labelwidth}{#1.}\sloppy}}{\end{list}}

\newcounter{itemlistc}
\newcounter{romanlistc}
\newcounter{alphlistc}
\newcounter{arabiclistc}

\newcommand{\fcaption}[1]{
        \refstepcounter{figure}
        \setbox\@tempboxa = \hbox{\tenrm Fig.~\thefigure. #1}
        \ifdim \wd\@tempboxa > 6in
           {\begin{center}
        \parbox{6in}{\tenrm\baselineskip=12pt Fig.~\thefigure. #1}
            \end{center}}
        \else
             {\begin{center}
             {\tenrm Fig.~\thefigure. #1}
              \end{center}}
        \fi}

\newcommand{\tcaption}[1]{
        \refstepcounter{table}
        \setbox\@tempboxa = \hbox{\tenrm Table~\thetable. #1}
        \ifdim \wd\@tempboxa > 6in
           {\begin{center}
        \parbox{6in}{\tenrm\baselineskip=12pt Table~\thetable. #1}
            \end{center}}
        \else
             {\begin{center}
             {\tenrm Table~\thetable. #1}
              \end{center}}
        \fi}

\def\@citex[#1]#2{\if@filesw\immediate\write\@auxout
	{\string\citation{#2}}\fi
\def\@citea{}\@cite{\@for\@citeb:=#2\do
	{\@citea\def\@citea{,}\@ifundefined
	{b@\@citeb}{{\bf ?}\@warning
	{Citation `\@citeb' on page \thepage \space undefined}}
	{\csname b@\@citeb\endcsname}}}{#1}}

\newif\if@cghi
\def\cite{\@cghitrue\@ifnextchar [{\@tempswatrue
	\@citex}{\@tempswafalse\@citex[]}}
\def\citelow{\@cghifalse\@ifnextchar [{\@tempswatrue
	\@citex}{\@tempswafalse\@citex[]}}
\def\@cite#1#2{{$\null^{#1}$\if@tempswa\typeout
	{IJCGA warning: optional citation argument
	ignored: `#2'} \fi}}

\def\fnt#1#2{\footnotetext{\kern-.3em
	{$^{\mbox{\sevenrm #1}}$}{#2}}}

 1
 1
 1

\font\tenbf=cmbx10
\font\tenrm=cmr10
\font\tenit=cmti10

\font\ninerm=cmr9

\begin{document}

\centerline{\tenbf A TECHNIQUE FOR DIRECTLY MEASURING}
\baselineskip=16pt
\centerline{\tenbf THE GRAVITATIONAL ACCELERATION OF ANTIHYDROGEN}
\vspace{0.8cm}
\centerline{\tenrm THOMAS J.~PHILLIPS}
\baselineskip=13pt
\centerline{\tenit Physics Department, Duke University }
\baselineskip=12pt
\centerline{\tenit Durham, NC 27708-0305 USA}
\vspace{0.9cm}
\abstracts{
The gravitational force on antimatter has never been directly measured.
A method is suggested for measuring the acceleration of antimatter
$(\overline g)$ by measuring
the deflection of a beam of neutral antihydrogen atoms in the Earth's
gravitational field.  While a simple
position measurement of the beam could be used, a more efficient measurement
can
be made using a transmission interferometer.  A 1\% measurement of
$\overline g$ should be possible from a beam of about 100,000 atoms, with the
ultimate accuracy being determined largely by the number of antihydrogen atoms
that can be produced.  A method is suggested for producing an antihydrogen beam
appropriate for this experiment.
}

\vfil
\rm\baselineskip=14pt
\section{Introduction}
\vspace{-0.7cm}
\subsection{Motivation}
\vspace{-0.35cm}
	There has never been a direct measurement of the gravitational
acceleration of antimatter.  Even the Witteborn and Fairbanks\cite{Fairbank}
experiment only measured the gravitational force on electrons and not
positrons.
So while  most physicists expect that antimatter will fall towards the earth
with the same acceleration as matter, we do not {\em know} that it will.
There are even some recent experimental results that suggest that the dark
matter problem is really a problem with gravity,\cite{MMN} and if this is true,
then the gravitational acceleration of antimatter will not be the same as
matter. Furthermore, it is likely that a theory could be found that would be
able to explain any experimental result.  It is simply a question of  how many
cherished assumptions would need to be thrown out in the  process.
Of course, even a result exactly as expected would be
important, and would provide an important constraint on theoretical attempts to
unify gravity with the rest of the Standard Model.

\vspace{-0.35cm}
\subsection{History}
\vspace{-0.35cm}

	The ability to make a direct measurement of the gravitational
acceleration of antimatter hinges on the availability of slow-moving
antimatter.  Until recently, what little antimatter has been
available on this planet has been relativistic.  In fact, it was the
lack of an appropriate source of positrons that prevented Witteborn
and Fairbanks\cite{Fairbank} from measuring the gravitational force on
positrons.  However, the development of electromagnetic trapping techniques
has made it possible to hold and cool particles (charged or neutral)%
\cite{Paul},  and the  construction of the Low Energy Antiproton Ring (LEAR) at
CERN has made it possible to trap substantial numbers of
antiprotons\cite{GG_cool_pbar}.  This has led to an experiment which will
attempt to measure the gravitational acceleration of antiprotons\cite{PS200}.
However, such a measurement is extremely difficult\cite{charged_fall},
since electromagnetic
forces on the charged antiproton can easily overpower the force of
gravity on the particle.  Nevertheless, by using $H^-$ ions as a control,
the experimenters hope to eventually make a 1\%
measurement of the gravitational acceleration of antiprotons in the earth's
gravitational field ($\gbar$).

	While a measurement of the gravitational acceleration of
antiprotons may be possible, a measurement made with neutral
antimatter would be free of the systematic problems which plague
a measurement made with charged particles.  A number of ideas
have been presented for ways to measure the earth's gravitational
force on antihydrogen\cite{GG_grav,Ital_grav,Sex_grav}.
However, most of these methods require
the ability to capture a substantial number of antihydrogen atoms in
traps.  Since producing antihydrogen and then capturing it in traps is
almost certainly much more difficult than producing antihydrogen
without catching it, a neutral antimatter gravity experiment that
does not rely on trapping neutral atoms should be easier to
perform.  Some designs for this kind of experiment are discussed
below.

\vspace{-0.35cm}
\section{Measuring the Deflection of a Beam}
\label{sec:fall}
\vspace{-0.35cm}

	The experiment suggested here is to measure the gravitational
deflection of a ``beam'' of neutral antihydrogen atoms.  The word ``beam'' is
in quotes because the intensity is unimportant; it is not necessary to ever
have more than one $\Hbar$ at a time.
However, the atoms would be produced with a velocity
component in a particular direction which is significantly larger than
the thermal velocity of the $\Hbar$, thus producing a ``beam''.
Since the antihydrogen beam is crucial to this experiment, and since it
is unlikely that sufficient quantities of antihydrogen could be produced to
make a beam by collimating an isotropic source, we will briefly discuss some
methods which might be used to produce the $\Hbar$ beam at the end of this
paper.

\begin{figure}
\begin{center}
\mbox{\epsfig{file=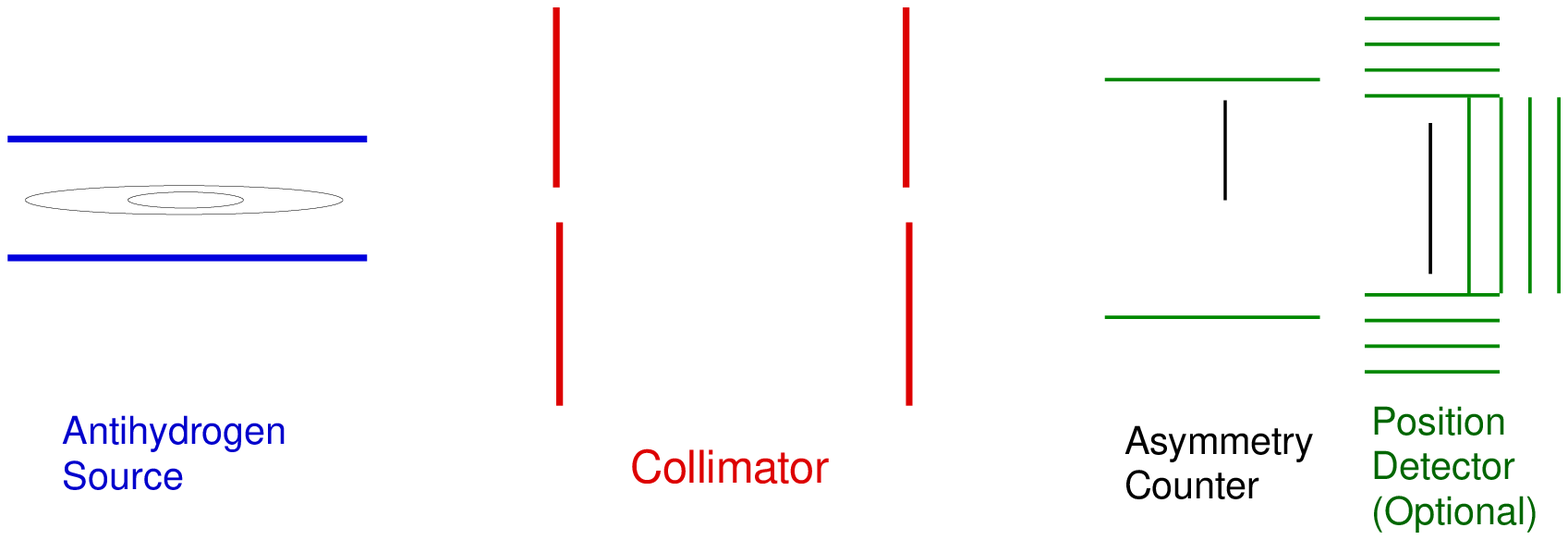,width=6in}}
\end{center}
\fcaption{Schematic of an experiment to measure
\protect{$\gbar$}
by directly measuring the deflection of a neutral antihydrogen beam.  Two
methods are shown for measuring the beam's position: a high precision vertex
detector; and a simple target covering half of the drift tube.  Scintillation
counters can be used to count the annihilations in this target and at the end
of the drift tube.  Not drawn to scale.
}
\label{fig:drift}
\end{figure}

	Perhaps the most obvious method to measure
$\gbar$ using the antihydrogen
beam would be to collimate the beam horizontally and measure its position
after it had propagated a sufficient distance in a drift tube.  A schematic of
such an experiment is shown in Figure~\ref{fig:drift}.  Two methods are shown
for measuring the position of the beam: a complex detector
for measuring the precise position
of each antihydrogen, and a much simpler method which only measures the
asymmetry.  In the first method, the position of each
antihydrogen atom is measured at the end of a drift tube, either by measuring
the trajectories of the charged particles produced by the $\overline pp$
annihilation when the $\Hbar$ strikes the target, or by measuring the impact
position directly, perhaps by detecting the nuclear recoil from the
annihilation.

	While a complicated detector would be able to give an accurate
measurement
of the impact position of each antihydrogen atom, its many electronic
channels are unnecessary when the measurement could be made with
only two channels.  This could be done by covering half of the drift tube
(either the upper half or the lower half) with a target, and measuring the
up/down asymmetry of the beam by counting the annihilations at this target
and the annihilations at a target covering the entire drift tube further
downstream.  Note that the targets need not be active.  External counters could
be used to detect the charged particles coming from the annihilations.

\begin{figure}
\begin{center}
\mbox{\epsfig{file=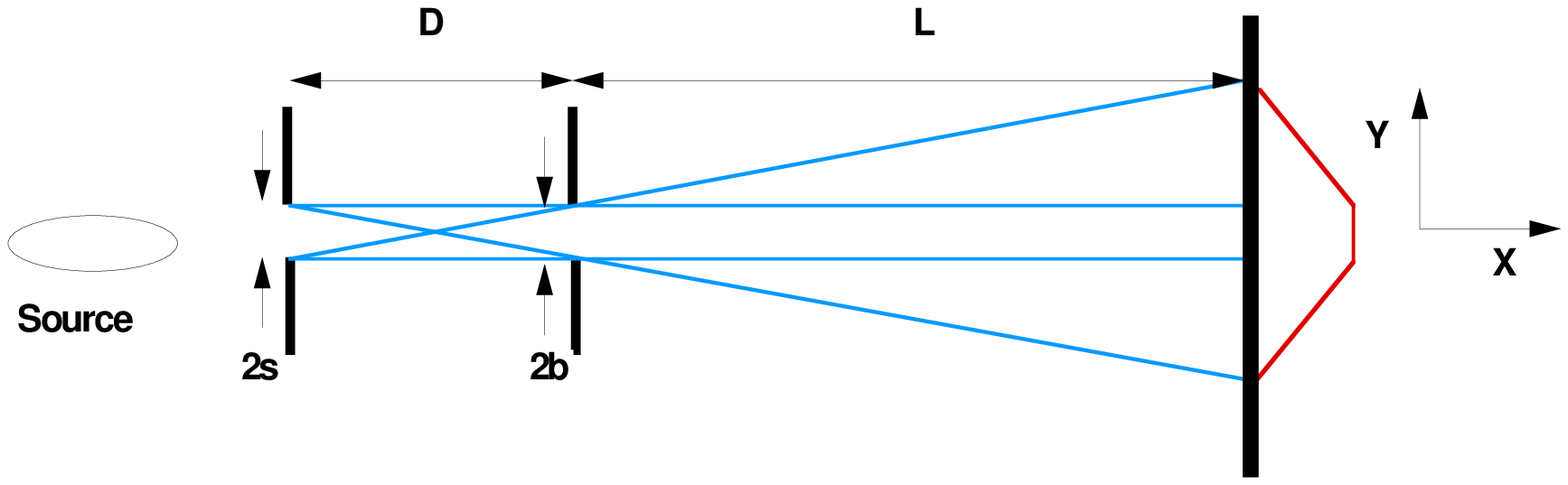,width=6in}}
\end{center}
\fcaption{Parameters characterizing a collimated beam. Not to scale.}
\label{fig:asymmetry}
\end{figure}

	The asymmetry $\alpha$ of a beam (for small deflections $\Delta y$)
is given by the formula
$$ \alpha \approx {\Delta y \over b} {D \over {D+L}} $$
where $2s$ and $2b$ are  the slit widths for the first and second slits of the
collimator, $D$ is the separation of the two slits, and $L$ is the drift
distance to the asymmetry measurement.  These dimensions are shown in
Figure~\ref{fig:asymmetry}.  The condition of small deflections is
that $\Delta y < b + (b-s) {L\over D}$.  The acceleration
of the beam between the collimator slits has been neglected in this
approximation.

	It is interesting to look at a concrete example to see how
well this method might work.
For the case where $b=s=2$ mm, the slits are separated by 1 meter, and the
asymmetry is measured 10 meters past the collimator ($D=1$ m and $L=10$ m), we
find a beam of particles travelling at $v_x = 500$ m/s would drop about 2~mm if
$\gbar=10$~m/s$^2$, and the asymmetry would be almost 10\%.

	The accuracy to which $\gbar$ can be measured is given by:
$$ {\sigma(\gbar) \over \gbar}
	= \sqrt{\left({\sigma(\alpha) \over \alpha} \right)^2
	+ \left( { 2 \sigma (v_x) \over v_x} \right)^2 }$$
where we have assumed that the dimensions of the apparatus are well known.
The uncertainty on $\alpha$ is statistical: $$\sigma(\alpha) = \sqrt{4 u d
\over N^3}$$ where $u$ and $d$ are the number of particles that are detected in
the upper and lower half, respectively, and $N = u + d$.  The uncertainty in
the velocity $\sigma(v_x)$ is difficult to estimate without knowing more
details about how the antihydrogen is produced.  A spread in the velocity is
not a problem so long as the velocity distribution is known.  However, for the
sake of this analysis, we can characterize the uncertainty in the velocity with
a temperature $T_u$:
$$ \sigma(v_x) = \sqrt{{kT_u}\over m}$$
where $k$ is Boltzmann's constant and $m$ is the antihydrogen mass.  We can use
this to calculate $\sigma(\gbar)$ in the previous example: for $N = 1000$ and
an
uncertainty in $v_x$ of $T_u = 1^\circ$K, $\sigma(\gbar)/\gbar = 0.5$, so
$\gbar$ could be distinguished from $-g$.
For $N=10^4$ and $T_u=0.1^\circ$ K, the
uncertainty in $\gbar$ would be about 16\%.

	The number $N$ in the above examples are the total number of
antihydrogen that make it through the collimator.  Since the antihydrogen is
produced with a finite transverse velocity distribution, only a fraction of the
the total antihydrogen produced will pass through both slits.  If the
transverse velocity is characterized by a temperature of $1^\circ$ K, then only
0.9\% of a 500 m/s beam will pass through a $\pm 2$mm slit at 1 meter, and only
half of this will pass through a second slit 1 meter past the first.  At
$T = 0.1^\circ$ K, the fraction is 1.4\% through both slits.  If we assume that
an additional half of the beam would be lost to horizontal collimation, then
a beam of about $1.5 \times 10^6$ antihydrogens is needed at $0.1^\circ$ K to
get $10^4$ through the collimation, or $4.4\times 10^5$ at $1^\circ$ K to get
1000 through.

\vspace{-0.35cm}
\section{Measuring $\gbar$ with an Interferometer}
\label{sec:interfere}
\vspace{-0.35cm}

	While measuring the deflection of a collimated beam of antihydrogen
atoms may be one of the most direct ways to measure the gravitational
acceleration of antimatter, a more efficient measurement can be made using an
interferometer.  The concept is to set up an interference pattern with a pair
of diffraction gratings, and to measure the phase position of the interference
pattern with a third grating.  The phase shift caused by gravitation can be
measured either by rotating the apparatus with respect to gravity, by changing
the length of the interferometer, or by
comparing the phase shifts for beams of different velocities.

\begin{figure}
\begin{center}
\mbox{\epsfig{file=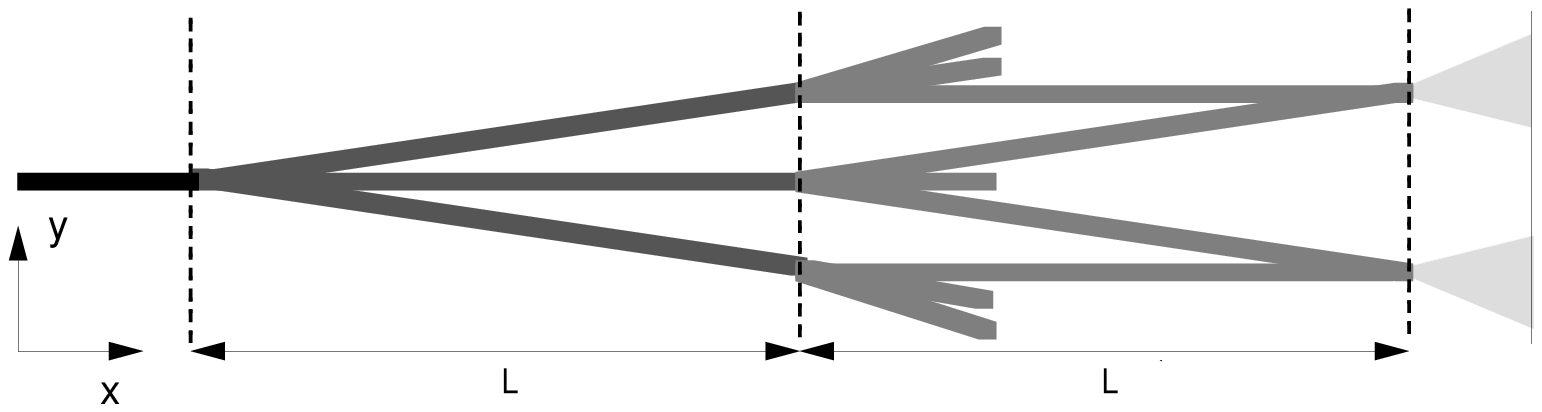,width=6in}}
\end{center}
\caption{Three-grating interferometer for measuring $g_{\rm antimatter}$.
The three diffraction orders shown will contain most of the transmitted
beam in roughly
equal amounts.  The orders which are drawn to the third grid cause an
interference pattern with a frequency that matches the grid spacing.  The
diffraction orders which are not followed to the third grid do not contribute
to this pattern, but rather cause a flat background.}
\label{fig:interferometer}
\end{figure}

	Perhaps the ideal interferometer for this experiment is a configuration
that has been used for both neutron and atom interferometers%
\cite{MIT_int,neutron_int}.  This interferometer consists of three
equally-spaced transmission gratings, each with identical grid spacing.  The
first two gratings set up an interference pattern that is independent of both
wavelength and the spacial coherence (incident wave direction) of the source%
\cite{interferometers}.  This interference pattern has a spacial frequency
which is equal to the grid spacing of the gratings, so the phase of the
interference pattern can be analyzed with a third identical grating.  The
interference pattern is localized in X (the direction perpendicular to the
grating planes), so while the distance between the first and second gratings is
arbitrary, the distance from the second to the third grating must match the
distance between the first and second gratings.  A diagram of the
interferometer is shown in Figure~\ref{fig:interferometer}.

	Not all of the diffraction orders from the first two gratings will
contribute to the interference pattern.  However, by using gratings with
roughly 50\% transmission (i.e. the slit width is half of the grating period),
the even diffraction orders are suppressed, and most of the transmitted
beam appears in
the $0^{th}$ and $\pm 1^{st}$ order in roughly equal amounts.  The orders that
will interfere are shown in Figure~\ref{fig:interferometer}.  Ideally,
approximately 4/9 of the beam transmitted through the second grating
will contribute to the interference pattern.

	The phase of the interference pattern can be measured by moving the
third grating in the \^{y} direction.
The transmission would be recorded as a function of the phase position
of the grating: the transmission is highest when the interference
peaks fall on the slits.

	The interference pattern ``falls'' by the same amount that
individual atoms fall
while transversing the interferometer.  Thus, for deflection $D$ given by
$$ D = {1\over2} \gbar t^2 = 2 \gbar {L^2\over v^2} $$
where $L$ is the separation between the gratings and $v$ is the velocity of the
antihydrogen, the phase shift $\Delta\phi$ is given by
$$\Delta\phi = D/d$$
 where $d$ is the line spacing (spatial frequency) of the grating.
It is important to note that while the interference pattern is independent of
velocity (wavelength), the deflection (or equivalently the phase shift) due to
gravity is not.  This means that a large velocity dispersion can wash out the
interference pattern when the phase shift due to gravity becomes significant,
so the beam used to make this measurement must either have a
sufficiently small velocity dispersion, or else the velocity of each
antihydrogen atom must be measured.  We will return to this point when we
discuss methods to produce the beam.

	By solving the above  formula for $\gbar$ we see that the uncertainty in
the measurement of $\gbar$ is given by
$${\sigma(\gbar)\over \gbar}
		= \sqrt{
		  \left( {\sigma(\Delta\phi)}\over\Delta\phi\right)^2
		+ \left( {2\sigma(v)}\over v\right)^2
		+ \left( {\sigma(d)}\over d \right)^2
		+ \left( {2\sigma(L)}\over L\right)^2 }. $$
If we assume that the distances $L$ and $d$ are well known, then the accuracy
of
the measurement is determined by how well the velocity is known, and how well
the phase shift is measured.  The latter term will be a function of the
statistics.

\vspace{-0.35cm}
\subsection{Example Interferometer}
\vspace{-0.35cm}

	An interferometer of the type described in the previous section
has been constructed by a group at MIT\cite{MIT_int} for use with a beam
of sodium atoms.  The parameters of the interferometer  are
$L=0.65$ m, $d=0.4\: \mu$m, and the slit width of each grating is half of the
period.  The sodium atoms were travelling at 1000 m/s, which corresponds to a
wavelength of 16 pm.  The interference signal had a contrast of 13\%, which was
about half of the calculated contrast (the difference was ascribed to a number
of small effects such as grating imperfections).  The authors were able to
measure the phase of the interference pattern to 0.1 radian with 4000 atoms
in the interference signal.

	While this particular interferometer configuration would not be ideal
for making the gravity measurement (for reasons that will be discussed below),
it can be used as an ``existence proof'' to see what kind of accuracies should
be possible with an interferometer.  As an example, let us consider an
antihydrogen
beam with velocity $v = 10^4$ m/s, which corresponds to a wavelength of 40 pm
and a deflection (for $\gbar=9.8$~m/s$^2$) of~0.8 $ \mu$m. An uncertainty in
the
phase measurement of 0.1 radian (for both a deflected and an undeflected beam)
would lead to an uncertainty in the measurement of $\gbar$ of 1\%.  In addition
to this, an uncertainty in the velocity corresponding to $T_u = 1^\circ$K would
lead to a total uncertainty of $\sigma(\gbar)/\gbar \approx 2\%$.

	The transmission through the third grating can be measured more
accurately for
antihydrogen than for sodium, since the antihydrogen which is not transmitted
annihilates on the grating and can be recorded.  Also, there should be little
or no background to the annihilations.  However, if we make the conservative
assumption that it still takes 4000 atoms in the interference pattern to
measure
the phase to 0.1 radian, then the total number of antihydrogen atoms needed in
the beam is $N_{\overline H} = 4000/\epsilon \times 2$.
The factor of two comes from measuring the phase of
both the deflected and the undeflected beam.  The efficiency $\epsilon$
includes two factors of 0.5 for transmission through the first and second
gratings, 4/9 for the diffraction orders which do
not interfere, and a factor of precisely $1/\pi$ for reality, which comes from
effects like the grating support structure, grating imperfections, and
misalignments.  These factors give
$\epsilon = 3.5\%$, not including any beam that might be lost if collimation is
necessary, for example if the gratings are too small to contain the entire
beam.
This means a beam of approximately $10^5$ antihydrogen atoms with well known
(or measurable) velocities could yield a 1\% measurement of $\gbar$ using
this interferometer.

	As mentioned above, the interferometer used for the sodium beam is not
optimal
for the antihydrogen measurement.  Because sodium has a much shorter wavelength
than antihydrogen (at the same velocity), and because the researchers wanted to
make a separated-beam interferometer, they were compelled to use a grating with
as small a period as possible.  Such a grating is very difficult to align:  the
period is too small to use lasers, and the alignment tolerances need to be less
than the period over the active area of the grating to maintain phase
coherence.  In addition, the total size of the grating cannot be made very
large, particularly since the grating's linearity must be better than the line
spacing over the entire active area.
A much better choice of the grating period for the gravity
measurement would be at least an order of magnitude larger.  This would greatly
simplify the construction and alignment of the gratings.  It would also allow a
laser to be used for alignment and to give an instantaneous phase
measurement.  The
interferometer could be lengthened to increase the gravitational deflection of
the beam and compensate for the increase in $d$ (and the corresponding decrease
in $\Delta\phi$).

\vspace{-0.35cm}
\section{Making an Antihydrogen Beam}
\vspace{-0.35cm}

	A number of authors have suggested methods for making
antihydrogen\cite{make_Hbar}.  The method which appears to be most
suitable for making a (nonrelativistic) beam uses overlapping ion traps
containing both antiprotons and positrons\cite{Gerry_make}.  Antihydrogen is
produced in the three-body process $\pbar + 2 e^+ \rightarrow \Hbar + e^+$.
The longitudinal motion of the antiprotons would be excited to make the
antihydrogen
form a beam.  The calculated production rate for this process is quite
high\cite{3bodyBfield}:
$$\Gamma = 6 \times 10^{-13} \left({4.2\over T}\right)^{9/2} n_e^2\: [s^{-1}]$$
per antiproton, where $n_e$ is the positron density in number per cm$^3$.
While this calculation includes the effect of the magnetic field on the motion
of the positrons, there are a number of uncertainties resulting
from other assumptions required for the calculation.  For example, the effect
of the magnetic field on the atomic states has been cited\cite{make_Hbar}
as potentially affecting the calculated rates.

\begin{figure}
\begin{center}
\mbox{\epsfig{file=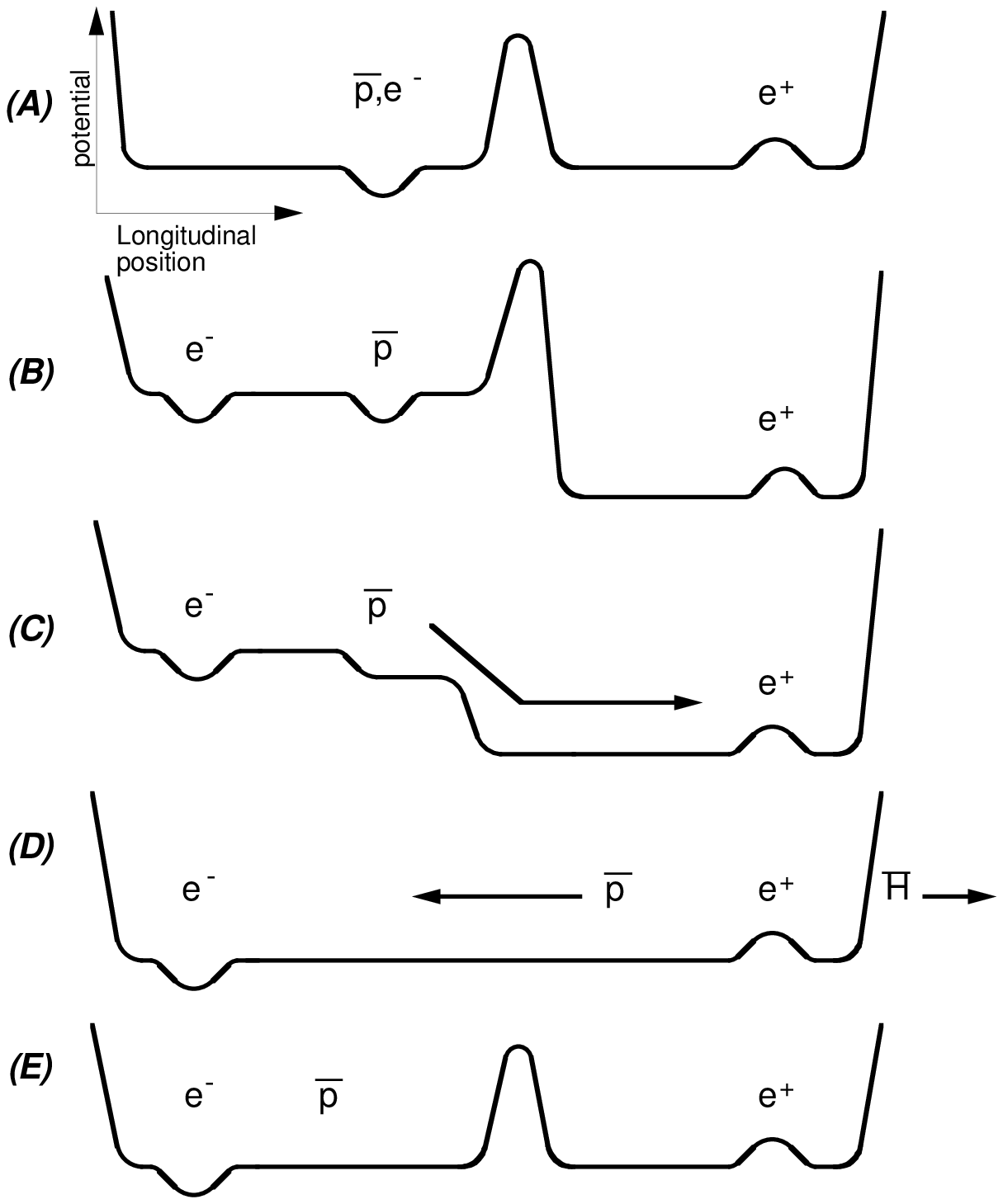,height=6in}}
\end{center}
\caption{Ion trap cycle for making a well-controlled antihydrogen beam.  In
(A),
the antiprotons are trapped and cooled with electrons on one side of the trap,
while positrons are loaded into an inverted well on the other side of the trap.
In (B), the electrons are ejected from the well containing the antiprotons and
saved for the next cycle in an adjacent well.  The potential of the antiproton
well is raised relative to the side of the trap containing the positrons.  In
(C), the barrier separating the antiprotons and the
positrons is lowered, and the
antiprotons are accelerated by the potential difference.  They pass through the
positrons and form some antihydrogen with a well-defined velocity and at a
well-defined time.  In (D) and (E), the returning antiprotons are again
trapped and cooled with the electrons, in preparation for another cycle.}
\label{fig:make_beam}
\end{figure}

	It is clear from discussions in sections  2 and 3
that the $\Hbar$ beam either needs
to have a narrow, well defined velocity distribution, or that the velocity of
individual antihydrogen atoms needs to be measured.  For the overlapping ion
traps, it might be possible to get a narrow velocity distribution if the
positrons occupy a fairly narrow region at the center of the harmonic well in
which the antiprotons are oscillating.  However, there could be problems from
transverse heating of the antiprotons, and there would be some longitudinal
cooling from collisions with the positrons.  In addition, the continuous
production of the antihydrogen would make it difficult to measure the velocity
of individual atoms.  Therefore, the more complicated ion trap shown in
figure~\ref{fig:make_beam} might be better for making the antihydrogen beam.

	The ion trap configuration shown in figure~\ref{fig:make_beam} could be
used to make an $\Hbar$ beam with a well-defined velocity and with a
well-defined time structure.  This would be done by alternately cooling the
antiprotons, and then accelerating them with a small potential difference
and passing them through a cold
positron plasma to form antihydrogen.  The trap and the interferometer could
be inside of a completely closed system at cryogenic temperature, so that
the plasmas could be held and cycled indefinitely.

	The rate equation given at the beginning of this section can be used to
estimate the number of antihydrogen atoms that would be produced per cycle.
For example, about 600 antihydrogens would be produced by $10^6$ antiprotons
with $v=1000$ m/s
passing through a 1 cm long plasma of positrons with a density of $10^7$ per
cm$^3$.  However, at much higher velocities, the assumptions that went into the
calculation begin to break down.  For example, the calculation assumes the ion
(antiproton) is at rest, which is a good assumption so long as the $\pbar$
velocity is significantly below the positron's thermal velocity (8000 m/s at
$4.2^\circ$K.  Also, the $\pbar$ needs to remain in the
plasma long enough for the atom to ``evolve'' through collisions to a low
enough energy state to survive without being field-ionized as they exit the
trap.  At $4.2^\circ$K, this evolution time is $8\times 10^{-7}$ seconds, or
about 0.8 mm at $10^3$ m/s.

	A major advantage of producing antihydrogen in this manner is that
the $\Hbar$ is formed at a specific time, so the velocity of the antihydrogen
can be measured by recording the annihilation time. (The time in the cycle when
the $\Hbar$ are formed can be determined by measuring
the difference in the average annihilation times at the different gratings.)
Another advantage is that the acceleration voltage can be changed
cycle-by-cycle.  This would allow high velocity (small deflection) and low
velocity (large deflection) data to be accumulated on alternate cycles, which
would eliminate systematic errors in the measurement of $\Delta\phi$ resulting
from drifts in the interferometer.  Another advantage of the grating
interferometer is that it does not care what atomic state the antihydrogen is
in, so long as it is neutral.  Most neutral atom traps, as well as an
interferometer produced by light\cite{Chebotayev} (which incidentally has the
potential to make very precise gravity measurements) generally require the
atoms to be in a specific atomic state.

\clearpage
\vspace{-0.35cm}
\section{Conclusions}
\nobreak
\vspace{-0.35cm}
\par\nobreak
	A direct experimental measurement of the gravitational acceleration
of antimatter has never been done, and the window of opportunity (the
availability of low energy antiprotons) to make this fundamental measurement
may be  about to close.  Several methods for making this measurement
using a beam of neutral antihydrogen were discussed; the best of these uses a
transmission-grating interferometer to measure the gravitationally-induced
phase shift.  This method should be capable of making a 1\% measurement with a
relatively small number ($10^5$) of antihydrogen atoms.

	A method was suggested for making a beam of antihydrogen with a small
velocity dispersion and with a definite time structure, which would allow the
velocity of individual antihydrogen atoms to be measured.
Of course,  no one has accomplished making  any non-relativistic antihydrogen
yet, so the antihydrogen production is the most speculative aspect of this
experiment.

	The ultimate precision of this method for measuring $\gbar$ (assuming
that a sufficiently large number of antihydrogen atoms can be produced) will be
determined by how well the $\Hbar$ velocity can be measured (a timing
measurement), and by how well the dimensions of the interferometer are known.
Even if the gravity measurement does not yield any surprises, the physics that
can be done with an interferometer is very rich (there have been whole
conferences devoted to neutron interferometry).  If the calculated $\Hbar$
production rate is correct and sufficient quantities of antihydrogen can be
produced, then very precise limits on the difference between $g$ and $\gbar$
could be set.
This is an experiment that should be pursued while the opportunity exists.

%

\end{document}